\documentclass[aps,amsmath,twocolumn,superscriptaddress,prl,showpacs]{revtex4}

\usepackage{graphicx}

\begin{document}

\preprint{}

\title{Engineering the spatial confinement of exciton-polaritons in semiconductors}

\author{R. Idrissi Kaitouni}
\affiliation{Institute of Quantum Electronics and Photonics, Ecole Polytechnique F\'ed\'erale de Lausanne EPFL, CH-1015 Lausanne, Switzerland}
\author{O. El Da\"if}
\affiliation{Institute of Quantum Electronics and Photonics, Ecole Polytechnique F\'ed\'erale de Lausanne EPFL, CH-1015 Lausanne, Switzerland}
\author{M. Richard}
\affiliation{Institute of Quantum Electronics and Photonics, Ecole Polytechnique F\'ed\'erale de Lausanne EPFL, CH-1015 Lausanne, Switzerland}
\author{P. Lugan}
\affiliation{Institute of Quantum Electronics and Photonics, Ecole Polytechnique F\'ed\'erale de Lausanne EPFL, CH-1015 Lausanne, Switzerland}
\affiliation{Institute of Theoretical Physics, Ecole Polytechnique F\'ed\'erale de Lausanne EPFL, CH-1015 Lausanne, Switzerland}
\author{A. Baas}
\affiliation{Institute of Quantum Electronics and Photonics, Ecole Polytechnique F\'ed\'erale de Lausanne EPFL, CH-1015 Lausanne, Switzerland}
\author{T. Guillet}
\affiliation{Groupe d'Etude des Semiconducteurs (GES), Universit\'e de Montpellier II, Place Eug\`ene Bataillon, F-34095 Montpellier, France}
\author{F. Morier-Genoud}
\affiliation{Institute of Quantum Electronics and Photonics, Ecole Polytechnique F\'ed\'erale de Lausanne EPFL, CH-1015 Lausanne, Switzerland}
\author{J. D. Gani\`ere}
\affiliation{Institute of Quantum Electronics and Photonics, Ecole Polytechnique F\'ed\'erale de Lausanne EPFL, CH-1015 Lausanne, Switzerland}
\author{J. L. Staehli}
\affiliation{Institute of Quantum Electronics and Photonics, Ecole Polytechnique F\'ed\'erale de Lausanne EPFL, CH-1015 Lausanne, Switzerland}
\author{V. Savona}
\email[]{vincenzo.savona@epfl.ch}
\affiliation{Institute of Theoretical Physics, Ecole Polytechnique F\'ed\'erale de Lausanne EPFL, CH-1015 Lausanne, Switzerland}
\author{B. Deveaud}
\affiliation{Institute of Quantum Electronics and Photonics, Ecole Polytechnique F\'ed\'erale de Lausanne EPFL, CH-1015 Lausanne, Switzerland}

\date{\today}

\begin{abstract}
We demonstrate the spatial confinement of electronic excitations in a solid 
state system, within novel artificial structures that can be designed having 
arbitrary dimensionality and shape. The excitations under study are 
exciton-polaritons in a planar semiconductor microcavity. They are confined within a 
micron-sized region through lateral trapping of their photon component. 
Striking signatures of confined states of lower and upper polaritons are found 
in angle-resolved light emission spectra, where a discrete energy spectrum and 
broad angular patterns are present. A theoretical model supports unambiguously 
our observations. 
\end{abstract}

\maketitle

Most of the major advances in semiconductor physics and technology over the 
last thirty years have been obtained thanks to quantum confinement of 
elementary excitations along one, two, or three spatial 
dimensions,\cite{hess94,alivisatos96,bimberg99,hartmann00} and simultaneously 
to the improvement of their coupling to the electromagnetic field. In this 
context, quantum dots\cite{bimberg99} represent the prototypical system. They 
are often called ``macroatoms''\cite{biolatti00} as they allow quasi-zero-
dimensional confinement of electronic states and display a discrete spectrum 
of energy levels. The quantum dot fabrication technique is usually based on a 
spontaneous formation process producing dots randomly distributed within a 
restricted range of sizes and shapes.\cite{bimberg99} This in turn limits the 
control over the energy-level structure and makes single-dot applications a 
challenging task. 

As an alternative to electron-hole pairs in quantum dots, confined states of 
other kinds of excitations in solids can be engineered. To this purpose, 
two-dimensional polaritons in planar semiconductor 
microcavities\cite{weisbuch92,houdre94,savona95} (MCs) are particularly 
suited. In MCs, the photon part of the polariton is provided by the optical 
modes of a Fabry-Prot planar semiconductor resonator, which are resonant with 
the exciton level of an embedded semiconductor quantum well. The dependence of 
polariton energy on its in-plane momentum has a quadratic behaviour, reminding 
of a massive particle, with an effective mass typically of the order of $10^{-
5}$ times the free electron mass.\cite{savona96} It is remarkable that, given 
this peculiar energy-momentum dispersion, a sizeable spacing of energy levels 
is expected already when the confinement extends over a few microns -- a quite 
unique situation in a semiconductor artificial structure that makes 
fabrication, positioning and optical addressing much easier than for other 
nanostructured systems. Owing to their peculiar nature of weakly interacting 
bosonic quasiparticles, confined polaritons would be an optimal system for a 
wide range of fundamental and applied studies. Polariton parametric 
processes\cite{stevenson00,langbein04} could be exploited for producing 
confined polaritons in quantum states displaying nonclassical properties like 
quantum correlations and entanglement\cite{karr04,savasta05}. This, joined to 
the ease of integration, optical manipulation and readout, could be the 
premise for a novel kind of quantum information device. Moreover, the discrete 
spectrum is the key feature\cite{lauwers03} to overcome the effect of quantum 
fluctuations that dominate a two-dimensional interacting Bose gas, and opens 
the way to the observation of collective many-body effects and long-range 
order.\cite{snoke02,savona05} In a very recent work,\cite{eldaif06} we have 
described a new paradigm of devices that should be able of producing laterally 
confined polariton states in a MC. A spectroscopical analysis has 
revealed a series of sharp emission lines that display avoided level crossing 
when varying the exciton-cavity detuning. In spite of this promising premise, 
however, a direct experimental evidence of the simultaneous spatial confinement of 
upper and lower polariton modes is still needed.

In this Letter we present conclusive direct evidence of polariton spatial 
confinement in a three-dimensional trap. The analysis is carried out by means 
of angle-resolved photoluminescence (PL) spectroscopy of both confined and 
extended polariton modes, and is supported by a theoretical model of the 
polariton states. The new technique for polariton lateral confinement consists 
in patterning a spatial region of slightly larger thickness on the top surface 
of the microcavity spacer layer, and growing the top Bragg mirror afterwards. 
To a larger thickness corresponds a lower 
frequency of the optical resonance. The spatial pattern then acts as a two-
dimensional confining potential for the photon mode. The barriers of this 
potential can be made very shallow by introducing small thickness variations. 
This gives rise to both spatially confined modes and a continuum of extended 
modes at energies above the barrier. As a consequence of the linear exciton-
photon coupling, the polaritons resulting from the strong coupling of these 
photon modes with the quantum well exciton, will also be characterized by a 
mixed spectrum containing both confined and extended states. In addition to an 
easier fabrication approach, this kind of structure presents considerable 
advantages with respect to micropillars,\cite{dasbach01} where photon 
confinement is obtained by etching the whole cavity body, thus resulting 
exclusively in confined modes. Furthermore, the quantum well and most of the 
photonic structure are not etched, thus preserving the quality factor of the 
original planar cavity.

\begin{figure}[ht]
\includegraphics[width=.47 \textwidth]{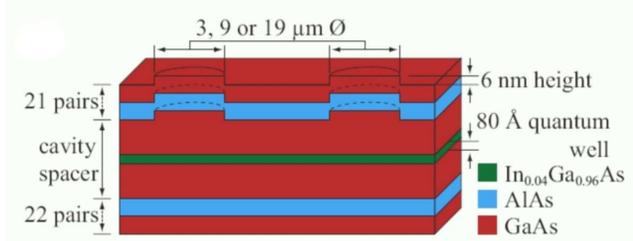}
\caption{Sketch of the sample (cross-section view). For clarity, 
the various lengths are not represented to scale.}
\label{fig1}
\end{figure}

The sample under investigation\cite{eldaif06} is sketched in Fig \ref{fig1}. 
It is a semiconductor MC consisting 
of a $\lambda$-thick GaAs spacer layer sandwiched between AlAs/GaAs 
distributed Bragg reflectors made of a 21 (top) and 22 (bottom) double 
$\lambda/4$-layers. Embedded at the MC center is a single 8 nm 
In$_{0.04}$Ga$_{0.96}$As quantum well, characterized by a sharp exciton 
resonance at 1.484 eV. A slight wedge of 2.4 meV/mm of the microcavity wafer 
allows varying the cavity detuning across the exciton resonance. Before 
growing the top mirror, a pattern of 6 nm height has been chemically etched on 
the cavity spacer using a photolithography mask. The pattern shape and height 
is preserved throughout the growth up to the Bragg mirror top surface. 
Circular mesas with nominal diameter of 3, 9 and 19 $\mu$m were patterned. The 
mesas are regularly spaced along the direction of the cavity wedge, so that 
mesas with varying exciton-cavity detuning for the laterally confined photon 
modes could be achieved. All our investigations were carried out at a 
temperature of 4 K, and consisted in PL measurements in the linear regime 
under pulsed off-resonance excitation at 760 nm, within the exciton continuum 
band. The sample was placed in the focal plane of a microscope objective. The 
excitation spot cross section had a Gaussian profile with 3 $\mu$m extension 
allowing investigation of a single mesa. The angular emission pattern is 
contained on the Fourier plane behind the objective, and is imaged onto the 
entrance slit of a monochromator. The slit selects a narrow stripe across the 
center of the Fourier plane, which is then dispersed inside the monocromator 
and detected by a CCD camera. In this way, thanks to the cylindrical symmetry, 
the spectral pattern of the emitted light, as a function of the energy and the 
emission angle, is directly displayed by the CCD.

\begin{figure}[h!]
\includegraphics[width=.49 \textwidth]{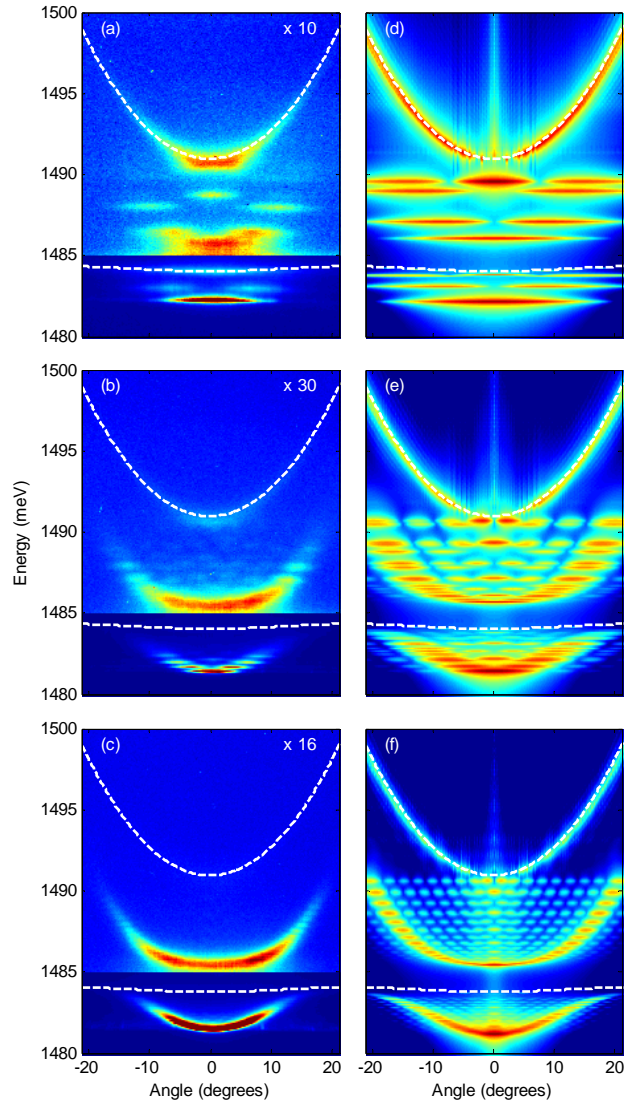}
\caption{Left: Measured polariton PL intensity (linear color scale from blue 
to red) as a function of energy and emission angle for the 3 $\mu$m (a), 9 
$\mu$m (b) and 19 $\mu$m (c) mesa. For clarity, the intensity above 1485 meV 
is multiplied by a constant factor, as indicated. Dashed: dispersion of the 
extended polariton modes, computed from a coupled oscillator model. Right: 
Intensity plot of the simulated polariton spectral density for the 3 $\mu$m 
(d), 9 $\mu$m (e) and 19 $\mu$m (f) mesa (color log-scale, 4 decades from blue 
to red).}
\label{fig2}
\end{figure}

A polariton state laterally confined in a small circular mesa is expected to 
emit light in a broad pattern of angular directions, which directly relates to 
the polariton wave function in momentum space. We take advantage of this 
feature for the characterization of lateral confinement. The measured vacuum 
field Rabi splitting, at unprocessed regions far away from the mesas, is 3.8 
meV, whereas the extrapolated linewidths are 200 and 500 $\mu$eV respectively 
for the bare cavity mode and for the exciton. Large patterned regions (250 
$\mu$m) fabricated for testing purposes\cite{eldaif06} reveal a similar polariton 
dispersion with a cavity mode energy 9 meV lower than in the unprocessed 
regions. This value is consistent with a 6 nm thickness variation from mesa to 
barrier, as indicated by a transfer matrix calculation of the microcavity 
resonance.\cite{savona95} When the excitation spot is focused on one mesa, the 
PL spectrum shows discrete narrow lines, in addition to a weak signature of 
the extended polariton states identical to that measured in the unprocessed 
regions of the sample. The dispersion pattern measured for 3, 9 and 19 $\mu$m 
diameter mesas are displayed in Fig. \ref{fig2} (a)-(c) respectively. The 
dispersion of the extended polariton (barely visible in the 19 $\mu$m mesa) is 
highlighted by dashed lines in the plots. From this part of the dispersion we 
can infer a positive detuning of 6.4 meV between the extended cavity mode and 
the exciton. The strongest spectral features in the three images appear below 
and above the lower extended polariton mode. In particular, the 3 $\mu$m mesa 
shows few discrete lines extending over broad angular regions, with energy 
spacings in the meV range. The 9 $\mu$m mesa shows a larger number of more 
closely spaced spectral lines, with a smaller angular spread. Finally, in the 
19 $\mu$m mesa these features approach a quasi continuous spectrum, while the 
angular spread is still smaller.

This general trend, consisting in a narrowing of the angular emission pattern at fixed energy 
and a decrease of the energy spacing as the mesa diameter increases, is 
observed for any value of the exciton-cavity detuning. Spatial confinement 
explains in a natural way these observations. Indeed, confinement induces a 
discrete energy spectrum and localization of the wave functions in real space, 
which in turn produces extended features in reciprocal space explaining the 
angular spread observed in the emission. The mesas therefore act like spatial 
traps. Are these confined states polaritons, namely linear superposition of 
exciton and cavity photons? For extended polaritons, the evidence of this 
strong coupling regime is usually given by the level anticrossing between the 
two branches in the energy-momentum dispersion. For confined states, the 
discrete energy spectrum makes level anticrossing more difficult to 
characterize. In the case of the 9 and 19 $\mu$m mesas in Fig. \ref{fig2} (b) 
and (c), however, the discrete levels form a pattern displaying a distinct 
level anticrossing at approximately 15 degrees, with a vacuum field Rabi 
splitting close to the 3.8 meV measured for the extended polaritons. This is a 
clear proof of strong coupling. For the 3 $\mu$m mesa, this feature is more 
difficult to characterize. However, we remark that in Fig. 2(a) the two levels 
at 1482.5 and 1483 meV, thus below the bare exciton energy, display the same 
angular pattern as the two levels lying above the bare exciton energy at 
1485.5 and 1486 meV. This clearly gives evidence to the fact that they are 
respectively lower and upper confined polariton states. This analysis proves 
that the strong coupling is preserved by spatial confinement and the species 
emitting are indeed mixed exciton-photon modes.

In order to support this interpretation, we compare the measured spectra to 
the prediction of a theoretical model.\cite{lugan06} The model consists in solving Maxwell 
equations inside the cavity, by making the assumption that the electromagnetic 
modes can be expressed as ${\bf E}({\bf r})={\bf E}({\pmb 
\rho})\exp(ik_z({\pmb\rho})z)$, namely by those of a locally planar 
Fabry-P\'erot resonator. This ansatz is justified by the small thickness variation 
and the large lateral extension of the mesas, as compared to the wavelength. 
The resulting photon modes are then included in a linear exciton-photon 
coupling Hamiltonian which is diagonalized. The mesas were assumed of circular 
shape and the nominal parameters of the samples were used for the 
calculations. Figs. \ref{fig2} (d)-(f) display the simulated polariton 
spectral density for the three different mesas. The energy-momentum structure 
of the simulated spectra should be compared to the experimental counterpart, 
whereas the relative spectral intensities in the PL data bear additional 
information on the polariton state populations, that cannot be accounted for 
in the simulated spectral density. A slight discrepancy in the energies of the 
smallest mesa is probably due to its shape not perfectly circular. In general 
however, the model faithfully reproduces both the energy position and angular 
extension of the various spectral features. This brings the final proof that 
the mesa structures are efficient traps for microcavity polaritons.

The intensity emitted from each polariton level is proportional, among other 
factors, to the number of polariton quasiparticles occupying that 
level.\cite{savona96} The measured data in Fig. \ref{fig2} (a)-(c) clearly 
show that the polariton population builds up in the lowest lying energy 
levels, indicating a rather effective energy- relaxation mechanism towards the 
bottom of the trap. A preliminary analysis indicates a Boltzmann-like 
distribution with a temperature of about $T=20$ K. This can be traced back to 
the presence of a large density of spatially extended states at energies above 
the confined states. In particular, in the data displayed in Fig. \ref{fig2} 
(a)-(c), the lower extended polariton branch (lower dashed line) is almost 
fully exciton-like, with a vanishing photon component that results in a very 
long radiative lifetime. These states act as a reservoir from which 
quasiparticles relax to the confined states at lower energy. The relaxation 
can take place through interaction with the thermal bath of 
phonons,\cite{tassone97} with free carriers,\cite{lagoudakis03} or via mutual 
polariton interaction.\cite{porras02} The broken translational symmetry of the 
confined system lifts the constraint of momentum conservation, thus enhancing 
the relaxation efficiency compared to the case of a planar microcavity.

\begin{figure}[h!]
\includegraphics[width=.50 \textwidth]{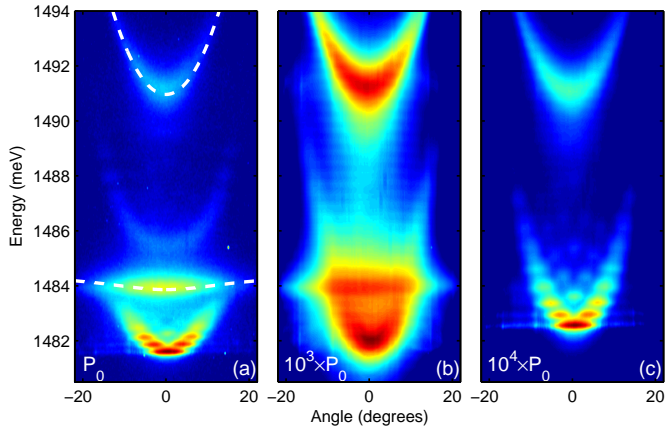}
\caption{(a) Measured PL intensity for the 9 $\mu$m mesa at low pump intensity 
$P_0$ in the linear regime (same data as in Fig. \ref{fig2}(b)). Dashed: 
extended polariton dispersion from a coupled oscillator model. (b) Pump 
intensity $10^3\times P_0$. (c) Pump intensity $10^4\times P_0$. Log-scale 
covering a factor 30 (a), 30 (b), and 100 (c) from blue to red.}
\label{fig3}
\end{figure}

We have investigated the behaviour of the system at increasing excitation 
intensity. In Fig. \ref{fig3} (a) we reproduce the PL intensity in the linear 
emission regime (pump intensity $P_0$) for the 9 $\mu$m mesa. The log-scale 
allows to better display the different features of both confined and extended 
polaritons. In Fig. \ref{fig3} (b) the pump intensity is 1000 times larger. 
Here, we observe a sizeable broadening of the spectral lines and the 
disappearing of strong coupling, displayed as a crossing (at $\pm 13$ degrees) 
between the bare exciton dispersion the cavity-like dispersion of the mesa 
modes. Both features are expected as a result of the density-dependent 
oscillator strength saturation and the collisional broadening of the exciton 
transition.\cite{khitrova99} At still higher pump intensity (Fig. \ref{fig3} 
(c)) the bound exciton spectral signature vanishes, the lasing threshold is 
reached and sharp emission lines through the bare electromagnetic modes of the 
mesa structure appear. We point out that the linear regime of strongly-coupled 
polaritons is preserved in this sample over two decades of pump intensity. No 
evidence of final-state stimulation with macroscopic occupation of the ground 
polariton level was observed, presumably due to the low saturation density of 
this single-well GaAs-based sample in which exciton bleaching dominates over 
polariton bosonic stimulation.

The physics of the present system profoundly differs from the recently 
achieved strong coupling of a single quantum dot in a 
nano-resonator.\cite{reithmaier04,yoshie04,peter05} In that case, the strong 
coupling is a direct consequence of the three-dimensional photon confinement, 
and produces a single pair of mixed two-level states. Here, we produce 
zero-dimensional trapping of polariton quasi-particles which are already in the 
strong coupling regime in the absence of the lateral trap. As a consequence, 
several confined and extended polariton states coexist which, due to their 
bosonic nature, can be occupied by more than one excitation quantum. All these 
features should help reaching the ideal situation of a weakly interacting cold 
Bose gas with a discrete energy spectrum, for which quantum collective 
phenomena are expected.\cite{lauwers03} On the other hand, the shallow 
confining potential makes it possible to design structures with two or more 
resonant traps having a significant tunnelling probability. This, together 
with the ease in resonant spatially-resolved optical excitation and detection, 
and to the preparation of nonclassical states via parametric polariton 
scattering\cite{savasta05} can lead to a variety of easily accessible schemes 
for coherent manipulation of the polariton quantum phase, thus opening the way 
to applications in quantum information technology.

In conclusion, we have succeeded in tailoring semiconductor microcavities in a 
way allowing to obtain for the first time spatial trapping of polaritons. An 
angle-resolved PL study demonstrates the high quality of the 
trapping as well as the simultaneous presence of extended states at higher 
energy, resulting in enhanced energy relaxation efficiency. Polariton traps 
have the unique property of displaying zero-dimensional confinement of an 
electronic degree of freedom in a solid, already on the micron scale. This 
guarantees ease of fabrication, high reproducibility and on demand access to a 
variety of spectral features (e.g. by tailoring the mesa shape and size), thus 
holding great promise for the realization of integrated solid-state quantum 
micro-devices.

This work was supported by the Quantum Photonics NCCR and project N. 
620-066060 of the Swiss National Research Foundation. We thank C. Ciuti, A. 
Quattropani and P. Schwendimann for helpful discussions. We are particularly 
grateful to W. Langbein for enlightening advice and H.-J. B\"uhlmann for 
technical help.

\bibliographystyle{PRSTY}

\end{document}